\documentclass{article}
\usepackage{arxiv}
\usepackage{graphicx}
\graphicspath{{Images_iThing-Self-Monitoring/}}
\usepackage[utf8]{inputenc}
\usepackage[T1]{fontenc}
\usepackage{hyperref}
\usepackage{blindtext}
\usepackage{mathtools}
\usepackage{relsize}
\usepackage{amssymb}
\usepackage{amsmath}
\usepackage{gensymb}
\usepackage{booktabs}
\usepackage{caption}
\usepackage{cite}
\usepackage{xcolor}
\usepackage{mhchem}
\usepackage{subcaption}
\usepackage{float}
\usepackage{subfloat}
\usepackage{tabularx}
\usepackage{listings}
\usepackage{algpseudocode}
\usepackage{algorithm}
\usepackage[numbers]{natbib}
\usepackage{makecell}
\usepackage{dirtytalk}
\usepackage{wrapfig}
\usepackage{picins}
\newcommand{\tabitem}{~~\llap{$\bullet$}~~}

\title{iThing: Designing Next-Generation Things with Battery Health Self-Monitoring Capabilities for Sustainable IoT in Smart Cities}

\author{ Aparna Sinha  \\
	Department of Electronics \& Communication Engineering\\
	IIIT Naya Raipur, Chhattisgarh, India\\
	\texttt{aparna@iiitnr.edu.in} \\
	\And
	Debanjan Das \\
	Department of Electronics \& Communication Engineering\\
	IIIT Naya Raipur, Chhattisgarh, India\\
	\texttt{debanjan@iiitnr.edu.in} \\
	\And
	Venkanna Udutalapally \\
	Department of Computer Science \& Engineering \\
	IIIT Naya Raipur, Chhattisgarh, India\\
	\texttt{venkannau@iiitnr.edu.in} \\
	\And
	Mukil Kumar Selvarajan \\
	Department of Instrumentation \& Control Engineering \\
	NIT Tiruchirappalli, Tamil Nadu, India \\
	\texttt{mukilkumar9@gmail.com} \\
	\And
	Saraju P. Mohanty \\
	Department of Computer Science \& Engineering \\
	University of North Texas, USA \\
	\texttt{smohanty@ieee.org} \\
}
\begin{document}

\maketitle

\begin{abstract}
An accurate and reliable technique for predicting Remaining Useful Life (RUL) for battery cells proves helpful in battery-operated IoT devices, especially in remotely operated sensor nodes. Data-driven methods have proved to be the most effective methods until now. These IoT devices have low computational capabilities to save costs, but Data-Driven battery health techniques often require a comparatively large amount of computational power to predict SOH and RUL due to most methods being feature-heavy. This issue calls for ways to predict RUL with the least amount of calculations and memory. This paper proposes an effective and novel peak extraction method to reduce computation and memory needs and provide accurate prediction methods using the least number of features while performing all calculations on-board. The model can self-sustain, requires minimal external interference, and hence operate remotely much longer. Experimental results prove the accuracy and reliability of this method. The Absolute Error (AE), Relative error (RE), and Root Mean Square Error (RMSE) are calculated to compare effectiveness. The training of the GPR model takes less than 2 seconds, and the correlation between SOH from peak extraction and RUL is 0.97.
\end{abstract}

%%%%%%%%%%%%%%%%%%%%%%%%%%%%%%%%%%%%%%%%
\section{Introduction}

\begin{figure*}[htbp]
        \centering
        \includegraphics[width=0.9\textwidth]{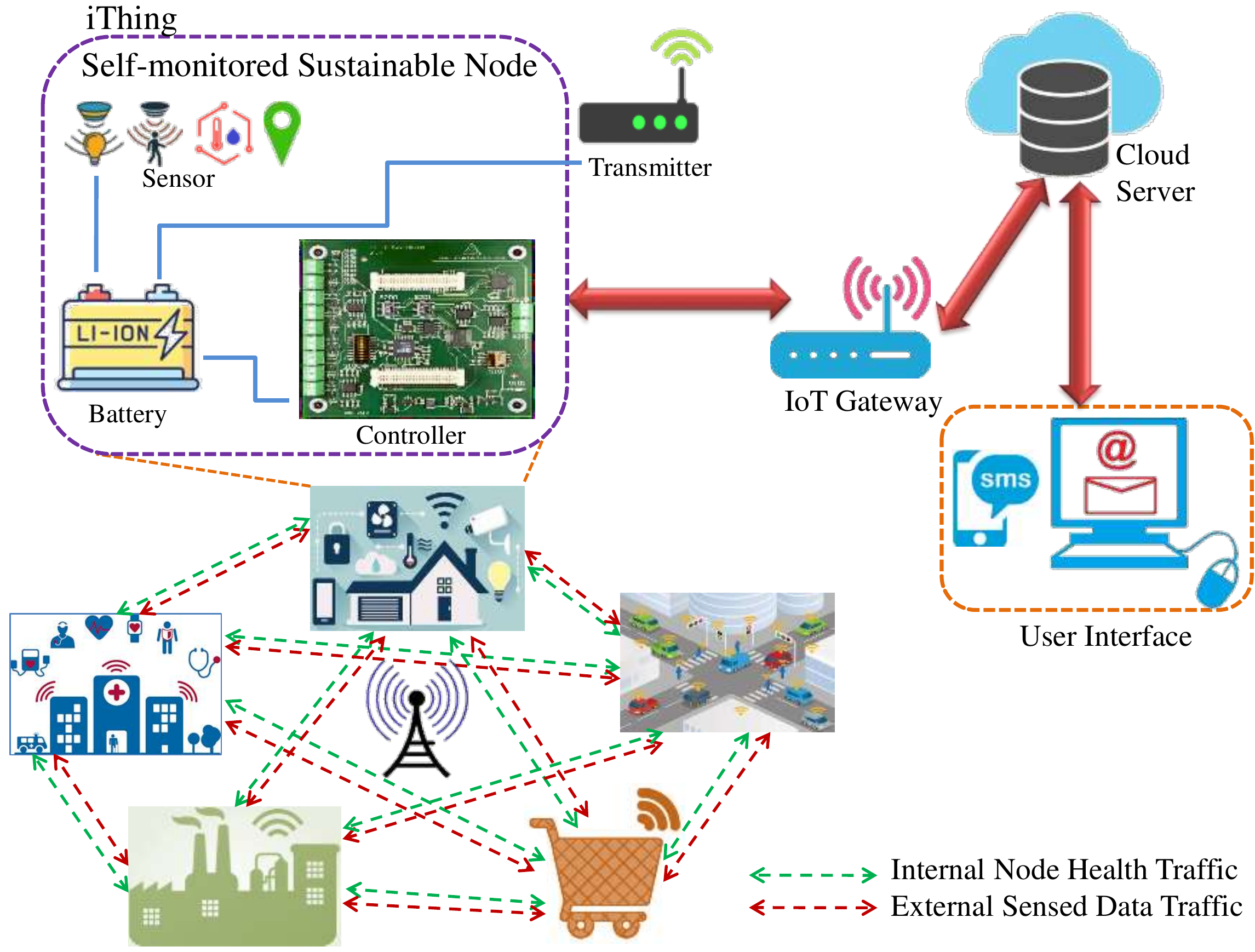}
        \captionsetup{justification=centering}
        \caption{Overview of the proposed resilient IoT-based solution for Smart City}
        \label{fig:overview}
    \end{figure*}

A smart city integrates technology with humanity for the well-being of its citizens and improvements of its resources \cite{Eternal-thing}, by the incorporation of the Internet of Things (IoT). In an IoT network, the sensor node is an important part, which has four main components - the sensor itself for data collection, the transceiver for data transfer to/from a local system or the Cloud, the microcontroller for controlling both the sensor and the transceiver, and the battery for powering the entire unit. The constant data collected and computed helps in various monitoring tasks in our environment. The real-world deployment of IoT has given rise to the challenge of monitoring the health and performance of the sensor nodes themselves \cite{zhang2007health}. If one or more components of sensor node malfunction, then the erroneous data may indicate a fault in the related system. Like any other component, sensors fail due to drift, bias, and precision degradation. Microprocessors usually last longer, but its memory \cite{Microprocessor_degradation_2} and other electronic components can fail with time. But in this paper, we will focus on battery health, as the death of the battery leads to the complete failure of the sensor node. For efficient and sustainable service, a resilient IoT-based solution has been thought of to communicate sensed data and health data, as shown in Fig. \ref{fig:overview}.

Efficient and sustainable computing for IoT has been the area of research interest, focusing on self-sustaining IoT nodes requiring the least possible maintenance in the long run \cite{Eternal-thing}. For IoT devices to work with the highest efficiency, energy usage must be accurately predicted \cite{Radio_power_practical_issues}. Energy Consumption can be reduced using a lightweight mutual authentication protocol for safe data transmission in smart cities \cite{Lightweight_authentication_IoT}, or using a more efficient A/D Converter for sampling sensor data \cite{ADC_efficiency}. Efficiency can also be improved by prioritizing more delay-sensitive machine-to-machine requests for a sustainable system \cite{sustainable_m2m}. All these developments have improved the IoT devices, but they have not talked about accurate and efficient ways to predict battery RUL, which is the main focus of this paper.

Rechargeable Lithium-ion batteries are extensively used in Electric Vehicles \cite{future_of_EV}, Mobile Phones and Laptops \cite{super_capacitors}, IoT devices, and many others as the power source due to their high energy density, low self-discharge, and prolonged lifespan compared to other battery types \cite{GPR}. Battery-powered IoT devices are mainly used in remote applications, whose uptime and functioning are critical. Since Li-ion batteries degrade over time due to formation of Solid Electrolyte Interphase film at the carbonaceous anode material \cite{safari2008multimodal}, the battery loses capacity and gains internal resistance. Hence, the in-built system should be able to accurately and reliably monitor the State of Health (SOH) and predict the Remaining Useful Life (RUL) of a battery to prevent over-discharge, overcharge \cite{intellbatt} and help in the timely replacement of the battery to avoid the failure of vital IoT instruments. In general, the SOH of a battery is the measure of the battery’s ability to store and deliver electrical energy. The RUL is defined as the remaining cycles or time before the battery reaches its End of Life (EOL), i.e., the time needed to reach 70\%\ or 80\%\ of the SOH. The SOH and RUL help in the Short-term and Long-term prediction of Battery life, respectively.\par
For Battery health prediction, data generation is also challenging, as Li-ion Battery has thousands of cycles, which causes reliability researchers to take several months or years before failure testing completes. The experimental methods \cite{noura2020review} and model-based methods \cite{ICA}\cite{kalman_fliter} are complex and unsuitable for IoT applications. Several Machine Learning models, such as Neural Network, Support Vector Machine (SVM), Gaussian Process Regression (GPR), have also been used for battery health prediction, but they were not developed considering remotely located IoT applications with low computational complexity. In most of these existing works, the raw data is transmitted to Cloud for health prediction, which increases the power needed by the transmitter. Since existing health prediction processes do not work for IoT devices deployed in the wild \cite{fafoutis2018predicting}, we have proposed a unique \textit{iThing} architecture capable of performing the SOH estimation and RUL prediction on-board using peak extraction method with minimal computational load and storage requirement.
    
The remaining paper is organized as follows: The related prior works and research gaps are mentioned in Section \ref{literature}. The contributions and novelty of the paper are stated in Section \ref{novelty}. The \textit{iThing} architecture, the peak extraction method, the methodology used, and the correlation analysis are explained in Section \ref{method}. The setup used for the experimental data extracted and the dataset used is discussed in Section \ref{experiment}. The accuracy and the efficiency of the novel peak extraction model proposed, and the prediction results are summed up in Section \ref{results}. Finally, the complete paper is summarized in Section \ref{conclusion}.

%%%%%%%%%%%%%%%%%%%%%%%%%%%%%%%%%%%%
\section{Related Works and Research Gap} 
\label{literature}

In general, there are three different methods for battery health estimation: Experimental, Model-based, and Data-driven methods. Experimental methods include measuring the direct Health Indicators (HIs) like internal resistance, impedance, and capacitance to estimate the battery SOH. The internal resistance of Li-ion batteries during charge-discharge cycles can be measured by Ohm's Law-based method \cite{wang2018internal}. Electrochemical Impedance Spectroscopy (EIS) may be used for Battery SOH derivation \cite{impedeance_based1}. Though accurate and straightforward, these experimental methods are laboratory-based, time-consuming, and need vast knowledge of the complex internal battery chemistry for precise prediction. Hence, they are not feasible for implementation in battery health estimation for IoT devices.   \par
Model-based methods, such as Kalman filter, least square-based methods, and simplified electrochemical models, are also used for battery health estimation. The influence of external resistance was investigated using the first-order equivalent battery model in \cite{resistance_based}. It is possible to estimate the battery SOH by Kalman filter and a Thevenin model \cite{kalman_fliter}, or by Incremental Capacity Analysis and Lorentzian function-based model \cite{ICA}. These methods are also accurate and robust, but have very complex mathematical structures. Hence, they are not feasible for low-power IoT applications with low computational power. \par
With the rapid development of Machine Learning and Artificial Intelligence, data-driven methods are becoming very popular, as they are non-parametric, do not consider the electrochemical principles much, do not require a deep understanding of the internal working of the battery, or complex mathematical equations for battery health prediction. A SOH prediction method utilizing Prior Knowledge-based Neural Networks and the Markov chain was proposed in \cite{PKNN}. Battery SOH may also be estimated using Relevance Vector Machine (RVM) and the Gray Model (GM) \cite{zhao2019hybrid}, or using SVM \cite{feng2019online}. Due to its several advantages, like lesser overfitting, fewer training data, etc., GPR has recently become more popular than other ML and AI techniques to estimate SOH and simultaneously train the RUL model \cite{GPR}. In this paper, the GPR technique predicts the RUL for maximum accuracy with a minimal computational load. \par

\begin{table*}[htbp]
            \centering
            \caption{Related works on sustainable computing for IoT}
            \label{tab:prior_work_TSUCS}
            %\begin{tabular}{ |>{\centering\arraybackslash}m{0.1\linewidth}|>{\centering\arraybackslash}m{0.2\linewidth}|>{\raggedright\arraybackslash}m{0.63\linewidth}| } 
		\begin{tabular}{p{2cm} p{4.5cm} p{7.7cm}}
            \toprule 
                \textbf{Author} & \textbf{Topic discussed} & \textbf{Salient features}\\
                \hline
                Ram \textit{et al}. \cite{Eternal-thing} & Self-sustaining IoT node &{
\tabitem Combined security, solar-energy harvesting, and aging detection in a unified system for self-sustaining SEHS
} \\
		\hline
               Luo \textit{et al}. \cite{Radio_power_practical_issues} & Practical issues in radio powered IoT & {
\tabitem  Proposed a model for very accurately describes the energy harvesting process and the power consumption of a sustainable IoT device
}\\
		\hline
               Li \textit{et al}. \cite{Lightweight_authentication_IoT} & Lightweight mutual authentication methods for IoT & {
\tabitem A lightweight mutual authentication protocol is proposed based on a novel public key encryption method to make IoT more secure and efficient at the same time
}\\
		\hline
               Huang \textit{et al}. \cite{sustainable_m2m} & Sustainable machine to machine communication & {
\tabitem An admission control model is used to split the priority of the requests made by the devices in order to make machine-to-machine communication more sustainable.
}\\
		\hline
               Klingensmith \textit{et al}. \cite{ADC_efficiency} & More efficient analog to digital converters & {
\tabitem A unique data acquisition technique using a non-uniform FTT algorithm is proposed to increase power efficiency for sampling data from sensors
}\\
		\hline
		Guha \textit{et al}. \cite{Resistance_capacitance} & Resistance and capacitance-based battery health prediction & {
\makecell[l]{\tabitem A capacitance and resistance-based particle filtering\\model is combined, and a fused degradation model is\\proposed to predict battery degradation. \\
\tabitem Model is complex and challenging to implement in a\\resource-strained IoT device}
}\\
		\hline
		Lyu \textit{et al}. \cite{EIS} & SOH prediction based on Electrochemical Impedance Spectroscopy (EIS)& {
\makecell[l]{\tabitem It uses a neural network for RUL prediction\\
\tabitem The proposed method is fast and economical\\
\tabitem Neural network prediction error increases as cycles\\increase}
}\\
		\hline
		Liu \textit{et al}. \cite{Electrochemical} & RUL prediction using electrochemical model of battery & {
\makecell[l]{\tabitem An electrochemical-based particle filter model is used\\to predict RUL\\
\tabitem State variables of the new PF algorithm are selected\\from battery’s health characteristics rather than meaning-\\less fitting coefficients\\
\tabitem The model was too complex to implement in low-\\power IoT devices}
} \\
		\hline
		Mo \textit{et al}. \cite{Kalman_filter} & RUL prediction using Kalman filter and improved particle filter & {
\makecell[l]{\tabitem A new particle filter method is used, which combines\\Kalman filter and particle swarm optimization\\
\tabitem The combined model is not affected much by noise}
} \\
		\hline
		Hu \textit{et al}. \cite{HI_extraction} & RUL prediction based on GPR model & {
\makecell[l]{\tabitem Uses a unique health indicator extraction method to\\train the GPR model for accurate RUL prediction\\
\tabitem It uses 12 different parameters to train the model; the\\battery must fully charge every time}
} \\
		\hline
		\textbf{Current paper} & \textbf{Peak Extraction method} & {
\makecell[l]{\tabitem Uses a unique peak extraction method that represents\\all factors causing battery degradation to train model\\
\tabitem Since only one variable is involved, it is efficient and\\suitable for low-powered and remote IoT applications}
} \\
                \bottomrule
            \end{tabular}
        \end{table*}

The summary of the related work with methodologies used along with a brief discussion are given in Table \ref{tab:prior_work_TSUCS}. Although all these discussed methods have their merits, none of them are made with IoT devices in mind, i.e., they have complex mathematical models, require high computational power, no correction for local mutations for individual battery cells, and other shortcomings mentioned previously. To overcome these shortcomings, a novel peak extraction method has been proposed to estimate SOH and predict RUL efficiently.

\section{Contributions of the Current Paper} \label{novelty}

\subsection{Problem Addressed in the Current Paper}

A smart city can improve the performance of resources and human life by the integration of intelligent sensors. These sensors need an uninterrupted power supply to monitor and transmit the required data continuously. The deployment of smart sensors in the wild often faces several problems, of which continuous power supply is one of the main issues. Most of the IoT sensor nodes use rechargeable Li-ion batteries as the source of power. However, the batteries suffer degradation with time. Although some methods exist to predict the remaining battery life, they involve complex calculations, more storage space and are not self-sustaining. Therefore, such methods are not suitable for deployment in the low-power IoT sensor nodes, which need to run for a long time without external interference. Hence, this paper has tried to address the challenge of Battery Health Self-monitoring for IoT nodes in Smart Cities.

\subsection{Solution Proposed in the Current Paper}

In this work, the battery's Charging Capacity is considered the Health Indicator of the Battery. The State of Charge (SoC) is extracted from the Charging Capacity by a simple formula. As the battery degrades, the Charge Capacity and the SoC values decrease. The State of Health (SOH) is taken as the collection of the peak values of SoC from each charge-discharge cycle of the battery. This process reduces the computational complexity of the method, making it suitable for IoT sensor nodes. \par
The GPR model is used to predict the Remaining Useful Life (RUL) of the battery. Since the GPR model can work well with a small training dataset, the storage requirement problem is also addressed.

    \subsection{Novelty of the Current Work}
         The main contributions of this paper are as follows:
         \begin{itemize}
        \item This paper proposes a straightforward method for battery health prediction, optimized for low-power IoT applications, using a novel technique of peak extraction. A small training dataset is used, which reduces the storage requirement.
	\item The proposed \textit{iThing} architecture enables the entire Battery life prediction to be done on-board. Only the predicted health data is transmitted instead of the raw data, reducing the transmitter power requirement and making it suitable for low-power IoT applications.
        \item The peak SoC of each charge-discharge cycle of the battery is extracted. This peak is then defined as the SOH of the battery cell and closely represents the change in internal factors of the battery, like resistance and impedance of the battery, but reduces the computational complexity of the process.
        \item The extracted SOH is then used to train a GPR model to predict the RUL, whose predicted values closely represent the actual values. The data from different battery cells are compared to establish the accuracy and robustness of the prediction model. 
        \end{itemize}

%%%%%%%%%%%%%%%%%%%%%%%%%%%%%%%%%%%%%%%%%%%%%%%%%  
\section{The Proposed Novel Next-Generation Sensors for Sustainable IoT} 
\label{method}

 \subsection{iThing Architecture}
	
\begin{figure}[b]
        \centering
        \includegraphics[width=0.8\textwidth]{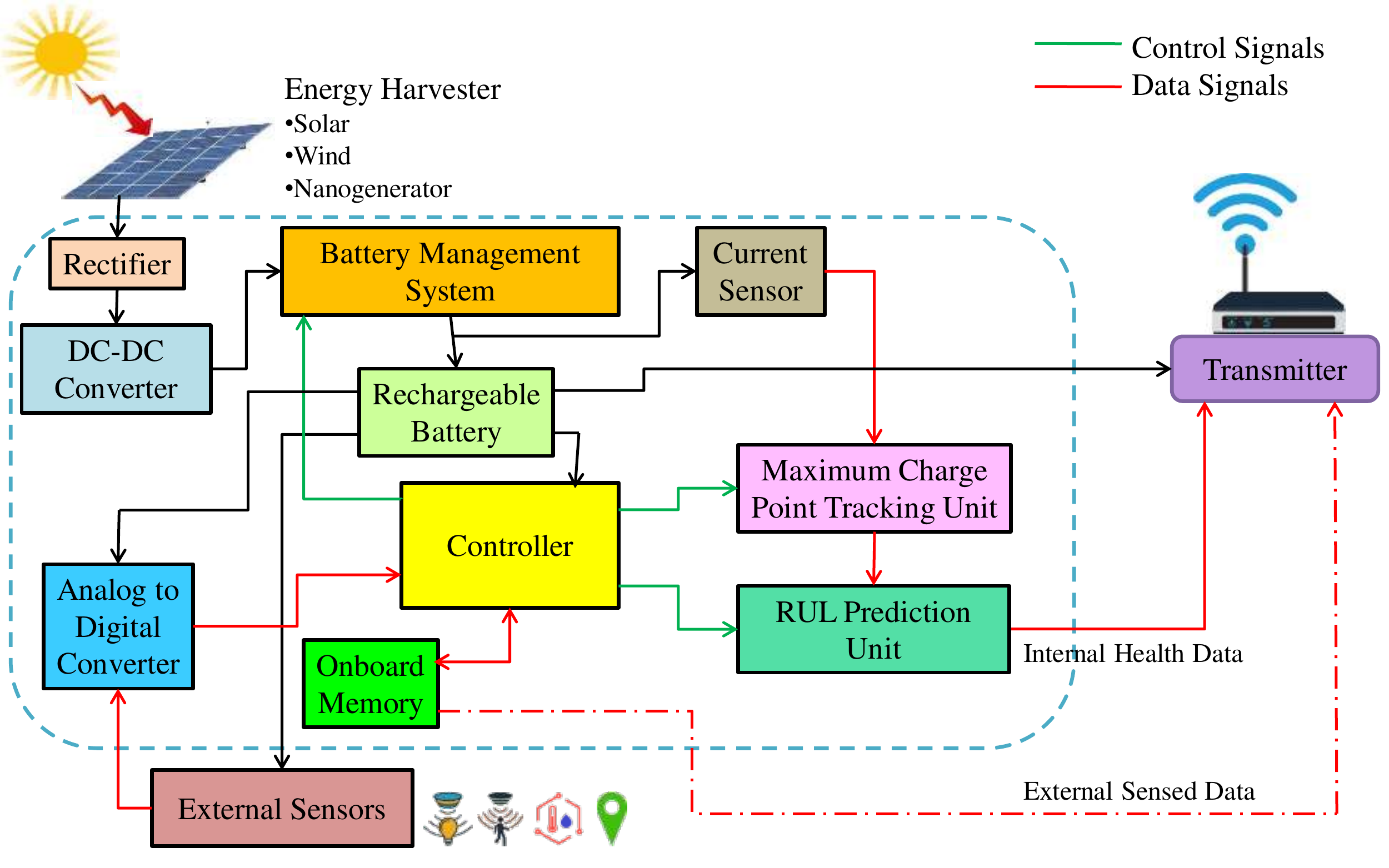}
        \captionsetup{justification=centering}
        \caption{Our Vision of iThing - Battery Health Self-monitoring for Sustainable IoT}
        \label{fig:architecture}
    \end{figure}

A unique hardware component called \textit{iThing} has been introduced for Sustainable IoT for Battery Health Self-monitoring in Sensor Nodes. The proposed architecture for \textit{iThing} is shown in Fig. \ref{fig:architecture}. The Battery is responsible for the power supply to the various components of the sensor node, like sensors, controller, and transmitter. The Battery Management System is responsible for recharging the battery in every cycle.

\begin{figure}[h]
        \centering
        \includegraphics[width=0.8\textwidth]{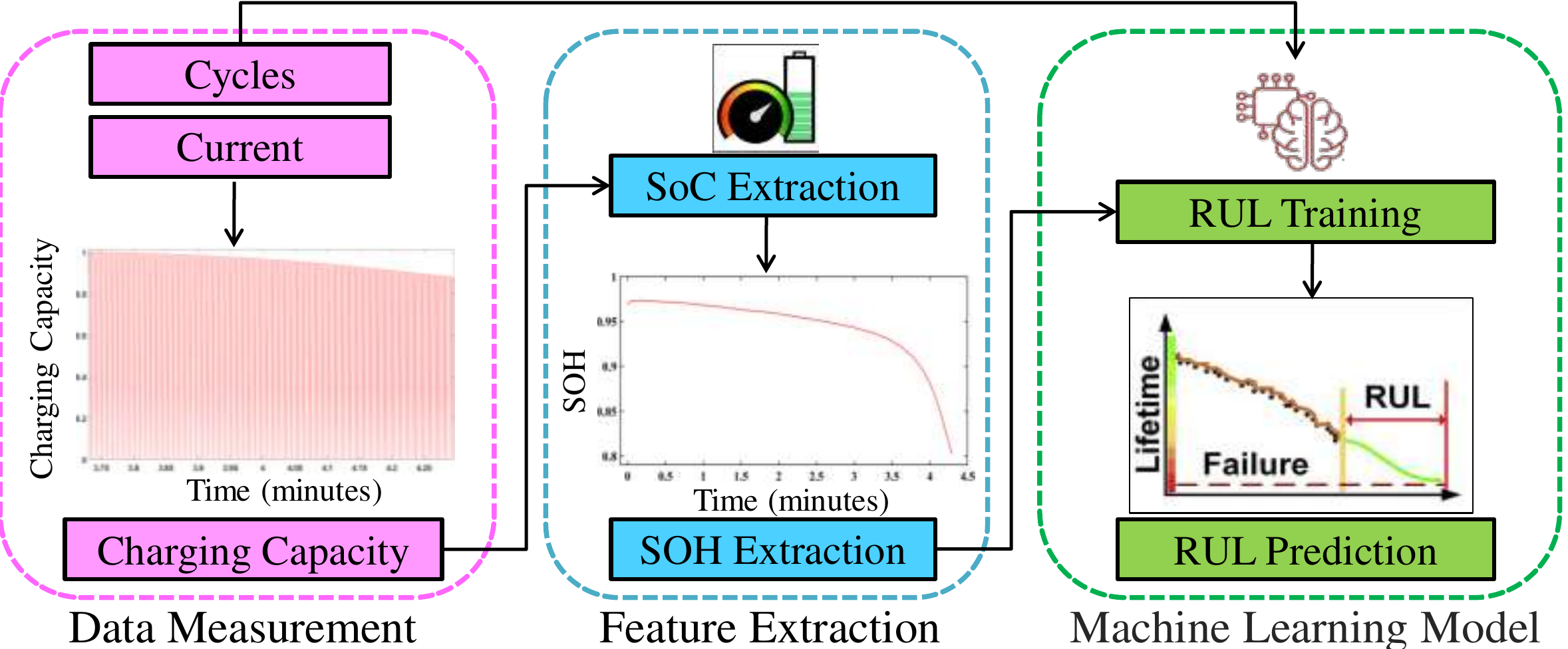}
        \captionsetup{justification=centering}
        \caption{A simple flowchart representing the process flow for RUL model training and prediction}
        \label{fig:flowchart}
    \end{figure}

The current sensor collects the time and current information and sends it to the Maximum Charge Point Tracking Unit, responsible for Charging Capacity calculation, SoC derivation, and SOH extraction. The extracted SOH data is then sent to the RUL Prediction Unit, where the GPR algorithm is used to predict the RUL. The entire calculation is done in the Sensor Node itself with minimal external interference, and the Internal Health Data is then transmitted to the Cloud. This process flow is shown in Fig. \ref{fig:flowchart}. The external sensors responsible for measuring the environmental parameters are also collected and sent for further processing and calculation to the Cloud.

In the following sub-sections, we will discuss the techniques used for the feature extraction method required to train the RUL prediction model.

\subsection{Extraction of SoC and SOH data from measured data}

\begin{figure}[h]
        \centering
        \includegraphics[width = 0.9\textwidth]{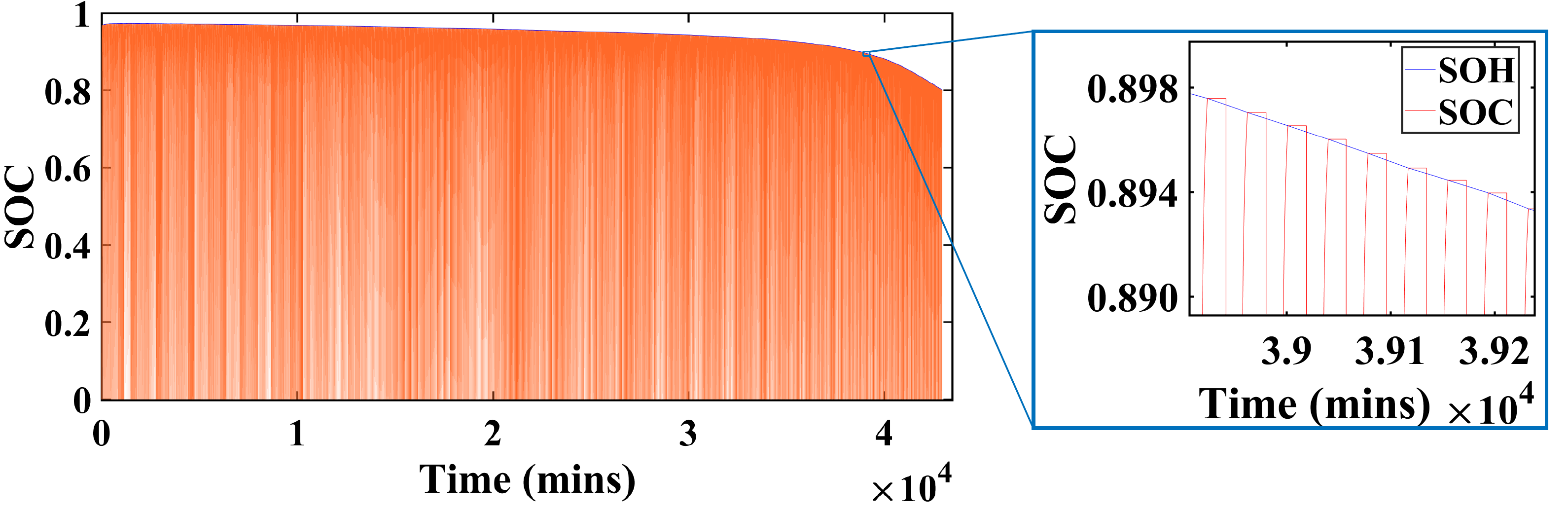}
        \caption{SOH from SOC by peak extraction}
        \label{fig:SOH_from_SOC}
    \end{figure}

        Since remotely operated IoT devices must operate at low power, the training model must not use too much computational power and memory. To reduce the computational complexity, the variables measured and the number of features used for model training are kept to two and one respectively, the variables measured is the charge capacity of the battery and the total time elapsed since the first charge-discharge cycle. The charge capacity of the battery is not directly calculated, it is calculated by integrating the current flowing into/out of the battery with respect to time. This can be done by either using a physical integrator circuit or computationally. However, the second method will result in errors due to approximation.
        \begin{equation}
            Charge\ capacity = \int{Idt}
        \end{equation}
        The measured charge capacity of the battery ($C_{curr}$) is divided by the battery’s nominal charge capacity ($C_{nomi}$) in Ah. The obtained value is considered as the SoC of the battery as given in equation \ref{SoC_eq}: 
        \begin{equation}
            SoC = \frac{C_{curr}}{C_{nomi}} \label{SoC_eq}
        \end{equation}
        When a battery undergoes multiple charge-discharge cycles, the maximum measured Charge Capacity of the battery at the end of the charging cycle invariably decreases compared to the previous cycle as the maximum charging capacity decreases. Due to the decrease in Charge Capacity, the calculated SoC value also decreases proportionally; this collection of peak SoCs is defined here as the SOH of the battery, as shown in fig. \ref{fig:SOH_from_SOC}.

    \subsection{Battery Life Prediction Algorithm} \label{GPR}

    Before we go into the RUL prediction processes itself, in this section, we will glance at the Gaussian Process Regression model used in this paper for training the RUL model. Several machine learning algorithms are popular; some of them include SVM, RVM, Na\"{i}ve-Bayes, and GPR. Out of all the models, GPR is more efficient and flexible, and accurate enough for RUL prediction than other models\cite{GPR}. Hence, we will be using the GPR method for RUL prediction. Below is the mathematics for the Gaussian Process Regression.\\
    Let us assume \textit{x} is the feature set (SOH) and \textit{y} is the target set (RUL) of the data. The regression model is defined as:
    \begin{equation}
        y = f(x) + \epsilon
    \end{equation}
    and
     \begin{equation}
         f(x) = x^Tw,
    \end{equation}
    where \(\epsilon\) represents the Gaussian noise of the model and $w$ represents the vector of weights of the regression model. The likelihood of the model can be defined as:
    \begin{equation}
        p(y|x,w) = \frac{1}{(2\pi\sigma^2_n)^\frac{n}{2}}\exp{(-\frac{1}{2\sigma^2_n|y-X^Tw|^2})}.
    \end{equation}
    Assuming the mean $\mu$ as 0, and the covariance matrix as \(\Sigma_p\) for the weights:
    \begin{equation}
        w \sim N(0,\Sigma_p).
    \end{equation}
    The posterior distribution is defined as:
    \begin{equation}
        p(w|y,X) = \frac{p(y|X,w)p(w)}{p(y|X)}.
    \end{equation}
    By writing the terms that are dependent on weights only (i.e.), prior and likelihood, we obtain:
    \begin{equation}
        p(w|x,y) \sim N(\Bar{w},A^{-1})
    \end{equation}
    where \(\Bar{w}\) is defined as \(\sigma^{-2}_n(\sigma^{-2}_nXX^T + \Sigma^{-1}_p)^{-1}Xy\). Since this posterior distribution is Gaussian with mean \(\Bar{w} = \frac{1}{\sigma^2_n}A^{-1}Xy\) and covariance \(A^{-1} = \sigma^{-2}_nXX^T + \Sigma^{-1}_p\). Let us assume that \(x_*\) is the SOH input whose RUL output \(y_*\) is to be predicted, then the Gaussian posterior can be written as:
    \begin{equation}
        p(f_*|x_*,X,y) = N(\frac{1}{\sigma^2_n}x_*^TA^{-1}Xy,A^{-1}x)
    \end{equation}
Here, the kernel function, which is nothing but the covariance function, is defined as:
    \begin{equation}
        k(x,x') = \phi(x)^T\Sigma_p\phi(x')
    \end{equation}
    In this paper, we will be using the squared-exponential function, also known as the radial basis function, as the kernel functions; it is defined as:
    \begin{equation}
        k(x,x') = \sigma^2\exp(-\frac{1}{2}(x-x')l^{-1}(x-x'))
    \end{equation}
    where \(\sigma^2\) is the correlation coefficient matrix, and \(l\) represents the characteristic length scale.\\
    The general pseudo-code of GPR is given in Algorithm I.
    
   \begin{algorithm}
    \caption{Battery Life Prediction Algorithm}
    \begin{algorithmic}[1]
    \Function{fit}{self, length\_scale=1, X, y}
        \State kernel\_ = rbf\_kernel(X, None, length\_scale)
        \State  lower = True
        \State  L = cholesky(kernel\_, lower = lower)
        \State self.alpha\_ = cho\_solve((L, lower), y)
        \State self.X\_train\_ = X
        \State self.L\_ = L
    \EndFunction

    \Function{predict}{self, length\_scale, X}
        \State  K\_star = rbf\_kernel(X, self.X\_train\_, length\_scale)
        \State  y\_mean = K\_star.dot(self.alpha\_)
        \State  lower = True
        \State  v = cho\_solve((self.L\_, lower), K\_star.T)
        \State  y\_cov = rbf\_kernel(X, None, length\_scale) - K\_star.dot(v)
        \State \textbf{return} y\_mean, y\_cov
    \EndFunction

    \Function{rbf\_kernel}{X, Y = None, length\_scale = 1}
        \If{Y is NULL} 
            \State dists = pdist(X / length\_scale)
            \State K = exp(-.5 * dists)
            \State K = squareform(K)
            \State fill\_diagonal(K, 1)
        \Else
            \State dists = cdist(X / length\_scale, Y / length\_scale, metric = \say{sqeuclidean})
            \State K = exp(-.5 * dists)
        \EndIf
       \State \Return{K}
    \EndFunction
    \end{algorithmic}
    \end{algorithm}
    
In the code mentioned in Algorithm I, the function \say{pdist} calculates the pair-wise standardized euclidean distance of the given matrix. The function \say{cdist} calculates the standardized euclidean distance between the two given matrices. The function \say{fill\_diagonal} is used to fill the diagonal of the given matrix with the value provided. The function \say{squareform} converts the square-form distance matrix to a distance vector. The function \say{cholesky} returns the cholesky decomposition of the matrix. The function \say{cho\_solve} solves the linear equation Ax = B when the cholesky decomposition of A is provided.
With this, we have discussed the Gaussian process fully.
   
    \subsection{Training of the RUL model}

Once the SOH is calculated, the collected data is used to train a GPR model to predict the RUL of the battery. Since this paper targets low-powered IoT devices, sampling is done at a low frequency to save computational costs. Also, since these batteries last for a long time due to less usage, lots of data is collected even at low sampling usage, making the ML model very accurate at prediction. Before beginning the model training, the practitioner fixed a SOH value to stop the model training. Hence, until the calculated SOH reaches the fixed setpoint, the model is continuously trained with the acquired data. Once the SOH setpoint is crossed, the model training is halted, and the prediction of RUL for the battery starts.
        
    \subsection{Correlation analysis}
    
The SOH extracted and RUL predicted from this model is dependent on measured values. Still, to quantify their dependency, we perform correlation analysis on the two variables (SOH and RUL) with the measured variable (Charge Capacity). This is necessary for analyzing the prediction performance of the GPR model. There are different correlation analysis methods like Spearman’s correlation coefficient, Kendall’s correlation coefficient, and Pearson’s product-moment correlation coefficient. Out of the aforementioned methods, only Pearson’s correlation \cite{correlationmath} coefficient accepts non-ordinal data. But Pearson’s coefficient calculation assumes that its data is linear, which is not the case in this paper. Hence, the data is divided, and a piecewise analysis is performed.
        
Let us assume \(x_i\) is the \(i^{th}\) feature of the feature set (SOH in this paper) with size n, and \(y_i\) is the \(i^{th}\) target of the target set (RUL in this paper) with size n. Hence, the Pearson's correlation coefficient is calculated using the following formula \cite{correlationmath}:
        \begin{equation}
            r = \frac{\Sigma_{i=1}^n(x_i-\Bar{x})(y_i-\Bar{y})}{\sqrt{\Sigma_{i=1}^n(x_i-\Bar{x})^2\Sigma_{i=1}^n(y_i-\Bar{y})^2}},
        \end{equation}
        where \(\Bar{x}\) is
        \begin{equation}
            \Bar{x} = \frac{1}{n}\Sigma_{i=1}^nx_i
        \end{equation}
        and \(\Bar{y}\) is
        \begin{equation}
            \Bar{y} = \frac{1}{n}\Sigma_{i=1}^ny_i
        \end{equation}

%%%%%%%%%%%%%%%%%%%%%%%%%%%%%%%%%%%%%%%%%%%%%%%%%%%%%%%%%%
\section{Experimental Study} 
\label{experiment}

\subsection{Dataset description}

In this section, we will discuss the dataset used for the battery aging test. This data set is drawn from experiments performed by Severson \textit{et al.}\cite{dataset} and it is as follows: Lithium Iron Phosphate (LFP)/Graphite cells manufactured by A123 Systems (APR18650M1A) are used.  These batteries were cycled in horizontal cylindrical fixtures on a 48-channel Arbin LBT potentiostat in a forced convection temperature chamber whose temperature was set at 30 \degree C. The battery specifications are given in Table \ref{tab:battery_specifications}.

\begin{table}[h]
        \centering
        \caption{Specifications of the battery}
        \label{tab:battery_specifications}
        \begin{tabular}{l|l} \toprule 
            \textbf{Parameter} & \textbf{Value}\\
            \hline
            Nominal Capacity & 1.1 Ah\\
            Nominal Voltage & 3.3 V\\
            Battery Cathode & \(\text{LiFePO}_4\)\\
            Battery Anode & Graphite\\
            \bottomrule
        \end{tabular}
\end{table}

\begin{figure}[h]
        \centering
        \includegraphics[width=0.9\textwidth]{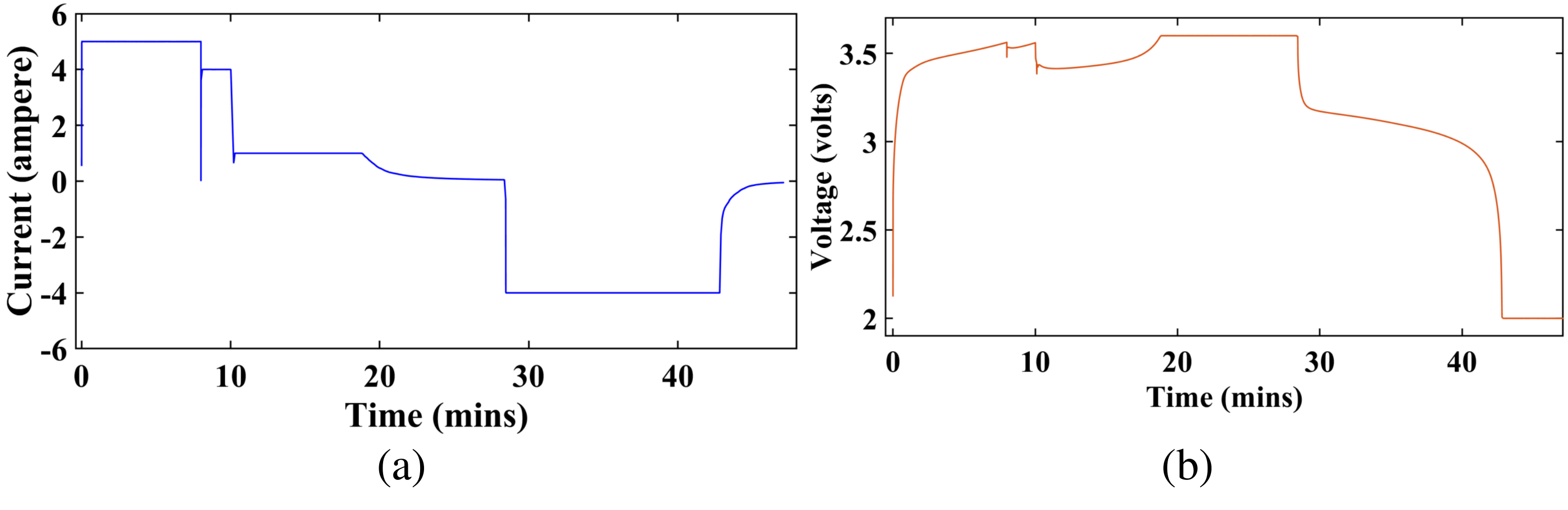}
        \captionsetup{justification=centering}
        \caption{The (a) Current versus time, and (b) Voltage versus time for one charge-discharge cycle for the dataset used}
        \label{fig:Dataset_graph_image}
    \end{figure}

The cells have been charged under a two-step fast charging condition. The battery is initially charged in Constant Current (CC) mode at 5 Columb until the cell reaches 67\% SoC, after that it is charged in CC mode at 4 Columb until the cell reaches 80\% SoC, charging time is fixed at 10 minutes for 0\% to 80\% SoC; after that, the battery is charged at 1 Columb Constant Current-Constant Voltage mode. The upper and lower cutoff potentials are 3.6 V and 2.0 V respectively, which are consistent with the manufacture’s specifications. These cutoff potentials are fixed for all current steps, including fast charging. After some cycling, the cells may hit the upper cutoff potential during fast charging, leading to significant constant-voltage charging. The battery is discharged at 4 Columb CC mode. The current and voltage graphs for one charge-discharge cycle are shown in Fig. \ref{fig:Dataset_graph_image}.
    
The computing platform used in the model training and correlation analysis is configured with Windows 10, 16 GB of RAM, and an AMD Ryzen 7 3750H CPU @ 2.3 GHz. The software used for training and testing the model was done using the scikit-learn library and the Pandas library in Python. The correlation analysis was done in R with the nlcor library. Data extraction and handling were done in Matlab.

 \subsection{Experimental Setup}
    
\begin{figure}[h]
        \centering
        \includegraphics[width=0.8\textwidth]{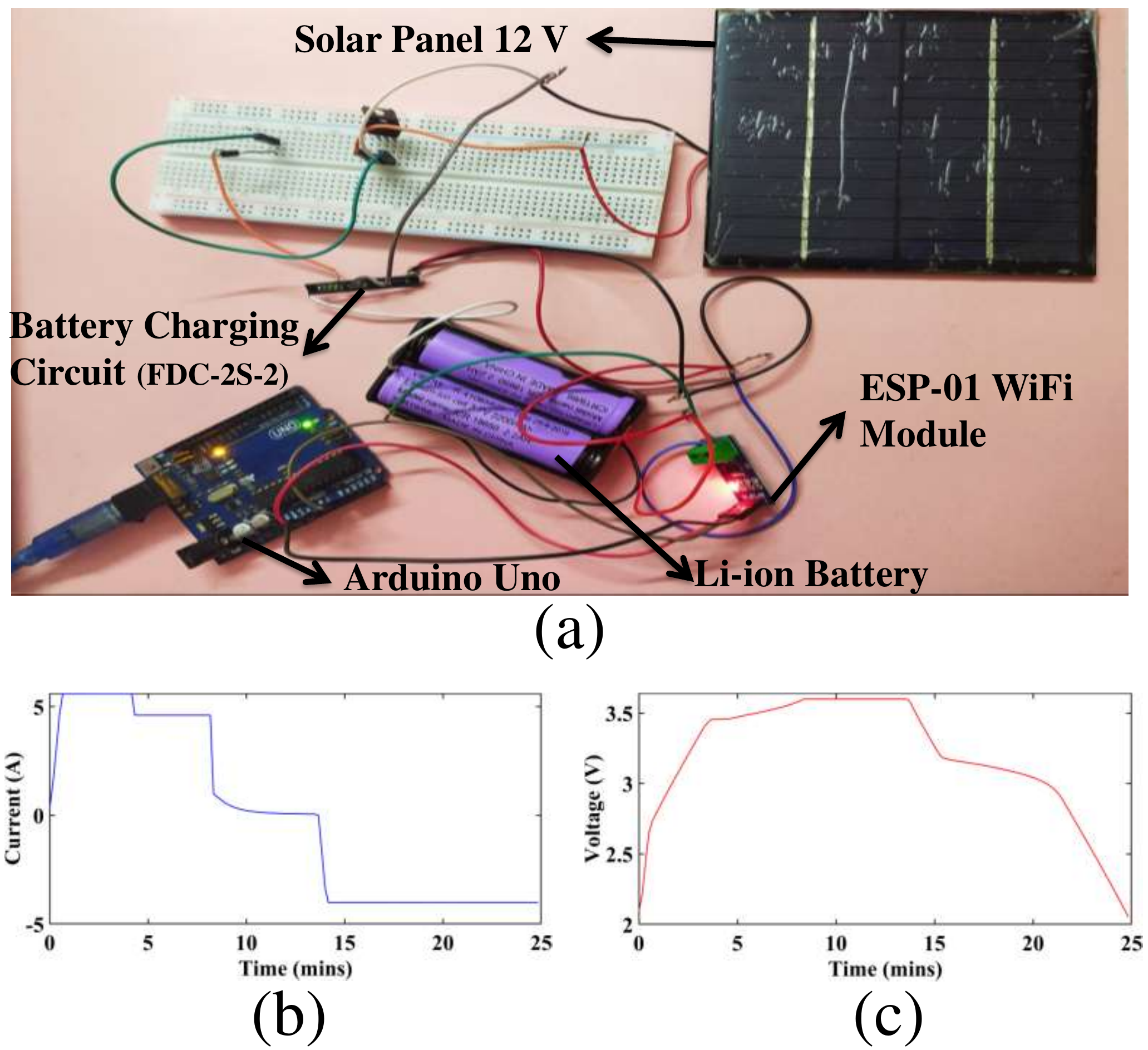}
        \captionsetup{justification=centering}
        \caption{The (a) Experimental Setup, (b) Current versus time, and (c) Voltage versus time for one charge-discharge cycle for the dataset used}
        \label{fig:exp_dataset_graph}
    \end{figure}

We have also extracted data for one cycle from a different battery, whose experimental setup, and the voltage and current versus time plots are depicted in Fig. \ref{fig:exp_dataset_graph}. 

The predicted RUL of the battery using the novel peak extraction model comes out to be 1141 cycles. The specifications of the battery used for this experiment are mentioned in Table \ref{tab:battery_specifications_exp}.

    \begin{table}[h]
        \centering
        \caption{Specifications of the battery used in the experimental setup}
        \label{tab:battery_specifications_exp}
        \begin{tabular}{l|l} \toprule 
            \textbf{Parameter} & \textbf{Value}\\
            \hline
            Nominal Capacity & 2.2 Ah\\
            Nominal Voltage & 3.7 V\\
            Model Number & ICR 18650\\
            \bottomrule
        \end{tabular}
    \end{table}

%%%%%%%%%%%%%%%%%%%%%%%%%%%%%%%%%%%%%%%%%%%%%%%%%%%%%%%
\section{Results and Discussion} \label{results}
    
    \subsection{RUL prediction}
    The battery cell RUL prediction model proposed in this paper predicts and provides the remaining cycles left of the battery life. To check and verify the accuracy and prediction performance of the trained model, three different error calculation methods are used in this paper. The types of error calculated are Absolute Error (AE), Relative Error (RE) and the Root Mean Square Error (RMSE).
    \begin{equation}
        AE = \vert RUL_{actual} - RUL_{predicted}\vert
    \end{equation}
    \begin{equation}
        RE = \frac{\vert RUL_{actual} - RUL_{predicted}\vert}{RUL_{actual}} * 100
    \end{equation}
    \begin{equation}
        RMSE =  \sqrt{\frac{\sum_{i=1}^n (RUL_{actual_i} - RUL_{predicted_i})^2}{n}}
    \end{equation}

\begin{figure}[h]
        \centering
        \includegraphics[width=0.9\textwidth]{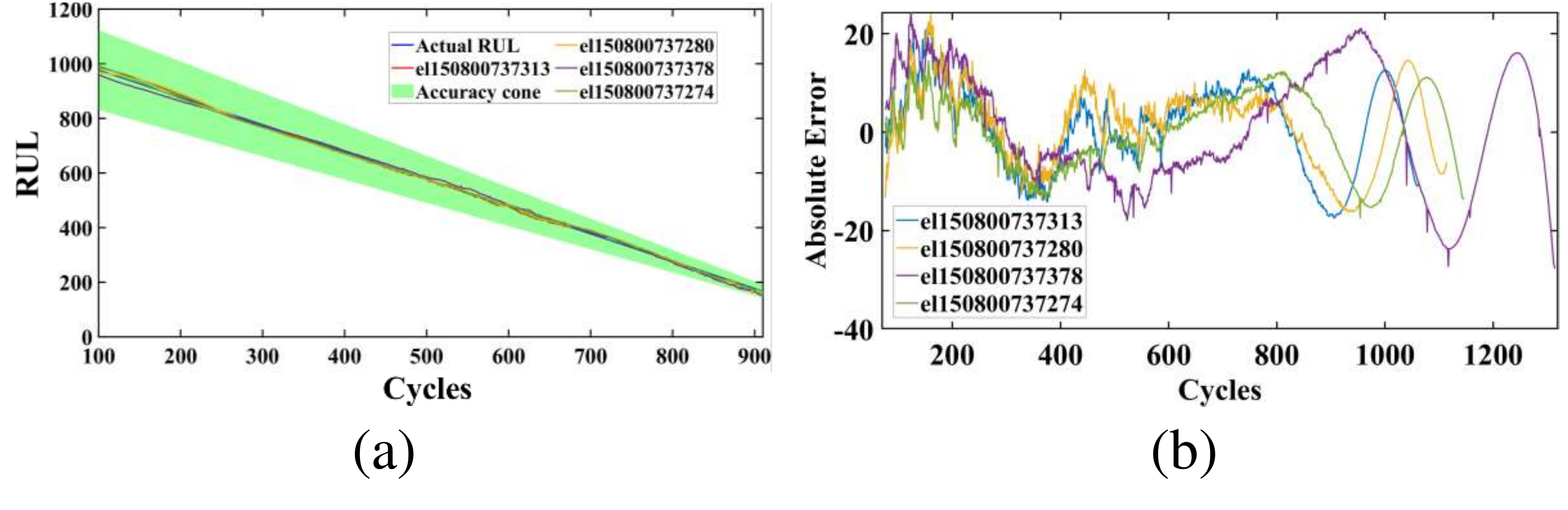}
        \captionsetup{justification=centering}
        \caption{These graphs show (a) the closeness of the actual RUL to the predicted RUL along with the PH margin, and (b) the Absolute Error in predicted RUL}
        \label{fig:RULact_vs_predicted}
    \end{figure}

In this paper, we focus more on computational efficiency rather than accuracy of the prediction model. The RUL prediction results are shown in Fig. \ref{fig:RULact_vs_predicted}. The exact values of the three errors for four different batteries are compared in Table \ref{tab:battery_compare}. 

 \begin{table}[h]
        \centering
        \caption{Comparison of results from different cells used in the dataset }
        \label{tab:battery_compare}
        \begin{tabular}{l|l|l|l} 
            \toprule 
            \textbf{Battery barcode} & \textbf{Maximum AE} & \textbf{Maximum RE} & \textbf{RMSE}  \\
            \hline
            el150800737313 & 22 cycles& 14.24\% & 9 cycles\\ 
            el150800737280 & 24 cycles& 15.04\% & 8 cycles\\
            el150800737378 & 28 cycles& 21.20\% & 12 cycles\\
            el150800737274 & 25 cycles& 20.93\% & 7 cycles\\
            \bottomrule
        \end{tabular}
    \end{table} 

\begin{figure}[h]
        \centering
        \includegraphics[width=0.9\textwidth]{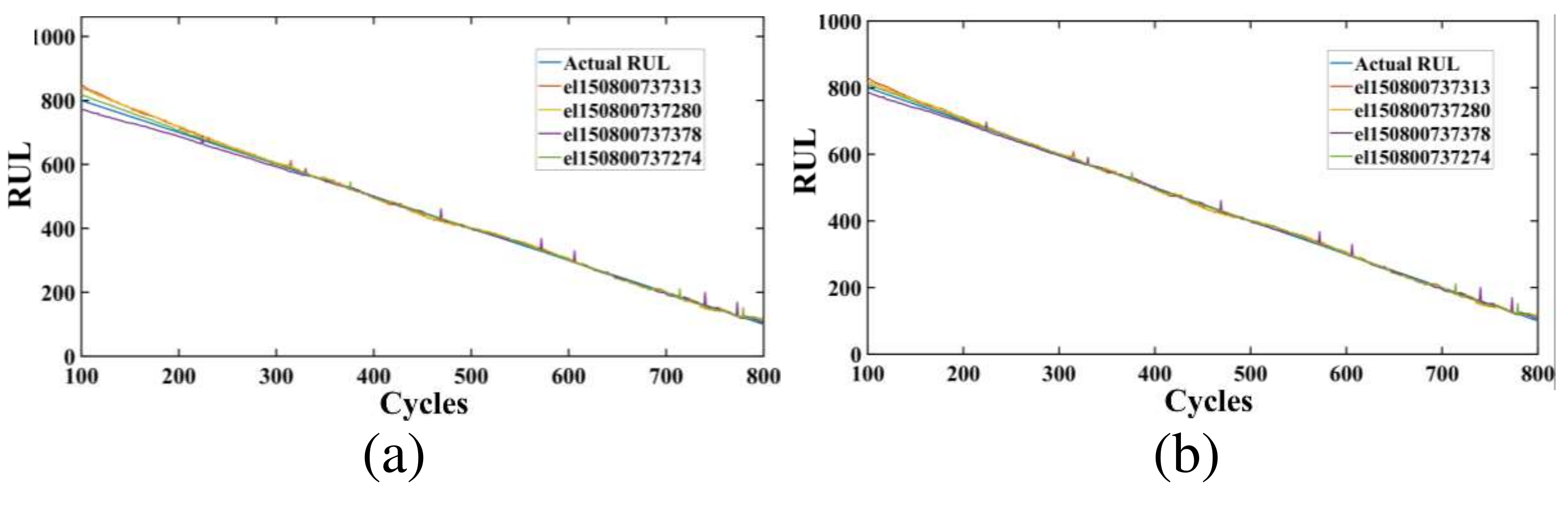}
        \captionsetup{justification=centering}
        \caption{The RUL prediction result of batteries starting from (a) 550$^{th}$ cycle, and (b) 600$^{th}$ cycle}
        \label{fig:err550_600}
\end{figure}

To verify the effectiveness, the proposed method is verified at different starting prediction points. As shown in Fig. \ref{fig:err550_600}, the starting prediction points are set as the 550$^{th}$ cycle and 600$^{th}$ cycle. The results from training the model are given in Table \ref{tab:battery_compare550} and Table \ref{tab:battery_compare600}, respectively. These results show that the proposed method can predict the RUL during the entire battery life, and the accuracy increases if the prediction starts at a later point.

 \begin{table}[h]
        \centering
        \caption{Comparison of results for four batteries with model trained for first 550 cycles}
        \label{tab:battery_compare550}
        \begin{tabular}{c|c|>{\centering\arraybackslash}m{0.15\linewidth}|c|c} 
            \toprule 
            \textbf{Battery ID} & \textbf{Actual RUL} & \textbf{Predicted RUL} & \textbf{AE} & \textbf{RE}  \\
            \hline
            el150800737313& 157 cycles& 175 cycles& 18 cycles & 11.69\% \\ 
            el150800737280& 144 cycles& 162 cycles& 18 cycles & 12.68\% \\
            el150800737378& 223 cycles& 198 cycles& 25 cycles & 11.25\% \\
            el150800737274& 155 cycles& 161 cycles& 6 cycles & 4.04\% \\
            \bottomrule
        \end{tabular}
    \end{table}

    \begin{table}[h]
        \centering
        \caption{Comparison of results for four batteries with model trained for first 600 cycles}
        \label{tab:battery_compare600}
        \begin{tabular}{c|c|>{\centering\arraybackslash}m{0.15\linewidth}|c|c} 
            \toprule 
            \textbf{Battery ID} & \textbf{Actual RUL} & \textbf{Predicted RUL} & \textbf{AE} & \textbf{RE}  \\
            \hline
            el150800737313& 107 cycles& 117 cycles& 10 cycles & 8.99\% \\ 
            el150800737280& 94 cycles& 103 cycles& 9 cycles & 9.37\% \\
            el150800737378& 173 cycles& 159 cycles& 14 cycles & 8.08\% \\
            el150800737274& 105 cycles& 108 cycles& 3 cycles & 2.66\% \\
            \bottomrule
        \end{tabular}
    \end{table}

The Prognostic Horizon (PH) is also calculated to determine how fast the model\rq{s} predicted RUL reaches the correct RUL within the margin of error \cite{gou2020state}.
    \begin{equation}
        PH = Cycle_{EOL} - Cycle_{i}
    \end{equation}
where $Cycle_i$ is the cycle at which \(RUL_{actual}*(1 - \alpha) \leq RUL_{predicted} \leq RUL_{actual}*(1 + \alpha) \) and \(Cycle_{EOL}\) is the cycle at which the battery reaches it\rq{s} End of Life. Here \(\alpha\) is the error value from 0 to 1. As seen in Fig. \ref{fig:RULact_vs_predicted}, the proposed method can predict RUL consistently within the desired accuracy cone specified as \(\alpha\) (where \(\alpha\) = 0.1). The PH is used to determine how fast the RUL prediction model is; the higher the PH value, the faster the prediction model reaches the required error specifications. Hence, the higher the PH value, the better the prediction model. The PH for battery \say{el150800737280} comes out to be 641 cycles which means the model reaches the acceptable error quickly. This does not mean the error constantly remains under the required specification, but it is accurate none the less. These results show that the proposed method for RUL prediction is robust, straightforward, and accurate so that it can be used for battery health monitoring in the \textit{iThing} sensor node. This enables the \textit{iThing} to operate independently for a long time and ensures that no critical monitoring data is lost due to the death of the battery.

\subsection{Computational efficiency}
    
Since extracting peaks reduces the training data by a lot, only 0.14\% of the data was used in training the model. Since the GPR model can't be trained with all the collected data as the measured charge capacity is an oscillating data set, there is no point in storing all the collected data except for testing or diagnostic purposes. Hence, the SOH and RUL that have been extracted and stored for training the RUL prediction model come down to only 15 KB of data, which can be discarded after the model's training if necessary, saving even more space. The time taken for the GPR model to train is less than 2 seconds; this estimation was done in an off-the-shelf computing device with a CPU clock speed of 900 MHz and RAM capacity of 1GB.

\subsection{Correlation between RUL and SOH}

\begin{figure}[h]
        \centering
        \includegraphics[width=0.9\textwidth]{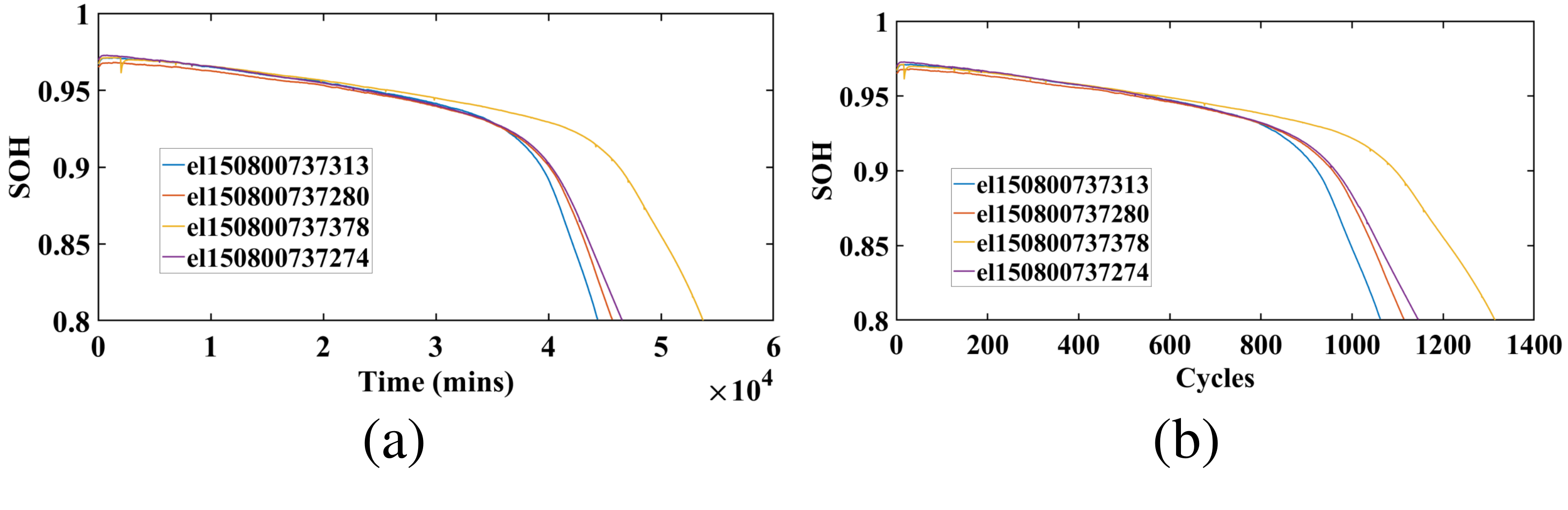}
        \captionsetup{justification=centering}
        \caption{These graphs represents the decrease in SOH with (a) time, and (b) number of cycles which represents the decrease in maximum charge capacity of the battery due to battery degradation}
        \label{fig:SOH_vs_time}
    \end{figure}

The correlation between RUL and SOH comes out to be 0.97, which means the calculated SOH is very closely related with RUL and hence the SOH closely represents the battery degradation, as seen in Fig. \ref{fig:SOH_vs_time}. This analysis was done in R using the \say{ncol} library.

Hence the error analysis results show that it is feasible for the model’s implementation in an IoT device including devices with low computational power and memory with acceptable levels of accuracy for general applications, this is also reflected in Table \ref{tab:result_compare} where the novel peak extraction technique is compared with other methods.

\begin{table}[h]
        \centering
        \caption{Comparison of results of different techniques}
        \label{tab:result_compare}
        \begin{tabular}{llll} 
            \toprule 
            \textbf{Method} & \textbf{Input Parameters} & \textbf{Output Parameters} & \textbf{Error} \\
            \hline
            Resistance \& Capacity-based\\models \cite{Resistance_capacitance}, 2018 & 2 & 1 & Average AE of 3 cycles \\ 
            EIS \cite{impedeance_based1}, 2019 & 5 & 1 & AE within 10\% of 240 cycles \\
            Electrochemical model \cite{Electrochemical}, 2020 & 5 & 1 & AE of 20 cycles maximum \\
            Kalman Filter \cite{kalman_fliter}, 2016 & 2 & 1 & Maximum AE of 7 cycles \\
            \textbf{Peak extraction} (proposed method) & 1 & 1 & RMSE error of 8 cycles \\
            \bottomrule
        \end{tabular}
    \end{table}

\subsection{Sustainability of the model}

As seen from the efficiency results, this RUL prediction model can be used for low powered applications, this allows remote and/or low-powered IoT nodes to operate and monitor their battery health remotely and can alert the user only for important actions like battery replacement. Since the RUL prediction is done locally, there is no need to transmit large amounts of training and testing data to the cloud for RUL prediction and hence can save up on energy costs. The device is integrated with clean and environmentally friendly power sources like solar cells to recharge the battery and power the device, this allows the device to operate for a prolonged period of time without external interference. These properties are compared with other papers in Table \ref{tab:result_compare_2}.

 \begin{table}[h]
        \centering
        \caption{Comparison of different IoT nodes}
        \label{tab:result_compare_2}
        \begin{tabular}{>{\centering\arraybackslash}m{0.2\linewidth}|>{\centering\arraybackslash}m{0.25\linewidth}|>{\centering\arraybackslash}m{0.25\linewidth}|>{\centering\arraybackslash}m{0.2\linewidth}} 
            \toprule 
            \textbf{Concept that uses IoT} & \textbf{Self Sustaining} & \textbf{Fully integrated} & \textbf{Self Monitoring} \\
            \hline
            IoT based energy saving devices \cite{IoT_energy_saving} & The nodes are powered by an external power source & The data is sent form the node to a master controller and hence no data is processed locally & The node only monitors the power use of the device and not the node itself \\
	    Monitoring pollution using IoT nodes \cite{IoT_pollution_monitor} & The node is battery powered and will have to be replaced & The node has a microcontroller which is used to perform calculations and classify the air pollution levels & The sensor node only monitors itself passively by approximating input data to avoid sensor errors \\
            Water management using WSN \cite{WSN_water_management} & The sensor nodes are battery powered and have to be replaced from time to time & The sensor node dose not process data locally to control the system & The sensor node does not monitor it\rq{s} own health \\
	    Self sustaining IoT based greenhouse \cite{IoT_agriculture} & Greenhouse and IoT device is powered by renewable energy & The device is integrated including it\rq{s} solar power source & The device only monitors the greenhouse, it does not monitor itself \\
            \textbf{Peak extraction} (our proposed method) & \textbf{The device is integrated with green sources for power and can last long without external interference} & \textbf{The device performs all required calculations and connects with the cloud only for user interference} & \textbf{The device can monitor it\rq{s} own health and inform the user in case of battery placement} \\
            \bottomrule
        \end{tabular}
    \end{table}

%%%%%%%%%%%%%%%%%%%%%%%%%%%%%%%%%%%%%%%%%%%%%%%%%%%
\section{Conclusion and Future Scope} \label{conclusion}
    
It is highly important and necessary for the battery health prediction system to be able to predict the RUL for the battery accurately and efficiently without consuming too much computational capability of the IoT devices, this applies especially for low-powered and remotely operated IoT devices. A novel peak extraction method is proposed in this paper to counter these problems. The main conclusions are summarized as follows: (1) A novel and simple peak extraction method is used to estimate SOH efficiently. (2) A single feature GPR based prediction model for efficient RUL prediction with low computational burden, the Pearson's correlation coefficient calculation also confirms the correlation between the extracted SOH and the predicted RUL. (3) The accuracy of the model is estimated with the absolute error and RMSE and the error values come out to be 22 cycles and 8 cycles for absolute error and RMSE respectively. With this we confirm that the proposed model is accurate and efficient enough to be implemented in IoT devices and can benefit the user a lot by alerting them for timely and planned replacement of the battery cell to avoid any crucial failures. Although this paper provides a novel method for battery health estimation, there are still some areas for improvement for this model. For example, the temperature inconsistencies of the battery were not considered, and only one type of battery (\(LiFePO_4\)) was experimented with and used to train and test the peak extraction method, which can be considered for future work.

%%%%%%%%%%%%%%%%%%%%%%%%%%%%%%%%%%%%%%%%%%%%%%%%%%%
%\balance
% IEEEabrv,
%\bibliographystyle{IEEEtran}
%\bibliography{IEEEabrv,Bibliography_X-ray}
\bibliographystyle{unsrtnat}
\bibliography{Bibliography_iThing-Self-Monitoring}

%%%%%%%%%%%%%%%%%%%%%%%%%%%%%%%%%%%%%%%%%%%%%%%%%%%
\section*{Author Biographies}
%\subsection*{ }
\parpic{\includegraphics[width=1in,clip,keepaspectratio]{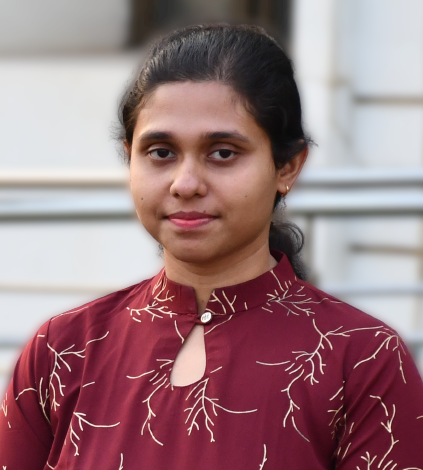}}
\noindent {\bf Aparna Sinha} (Student Member, IEEE) received the B.Tech. degree in Electronics and Communication Engineering from Techno India Saltlake, Kolkata, in 2013, and the M.Tech. degree in VLSI Design from Department of Radiophysics and Electronics, University of Calcutta, in 2020. She is currently working towards the Ph.D. degree from IIIT Naya Raipur, India. 
\\ Her current research interest includes Internet of Things (IoT) and Sensors. \par

\subsection*{ }
\parpic{\includegraphics[width=1in,clip,keepaspectratio]{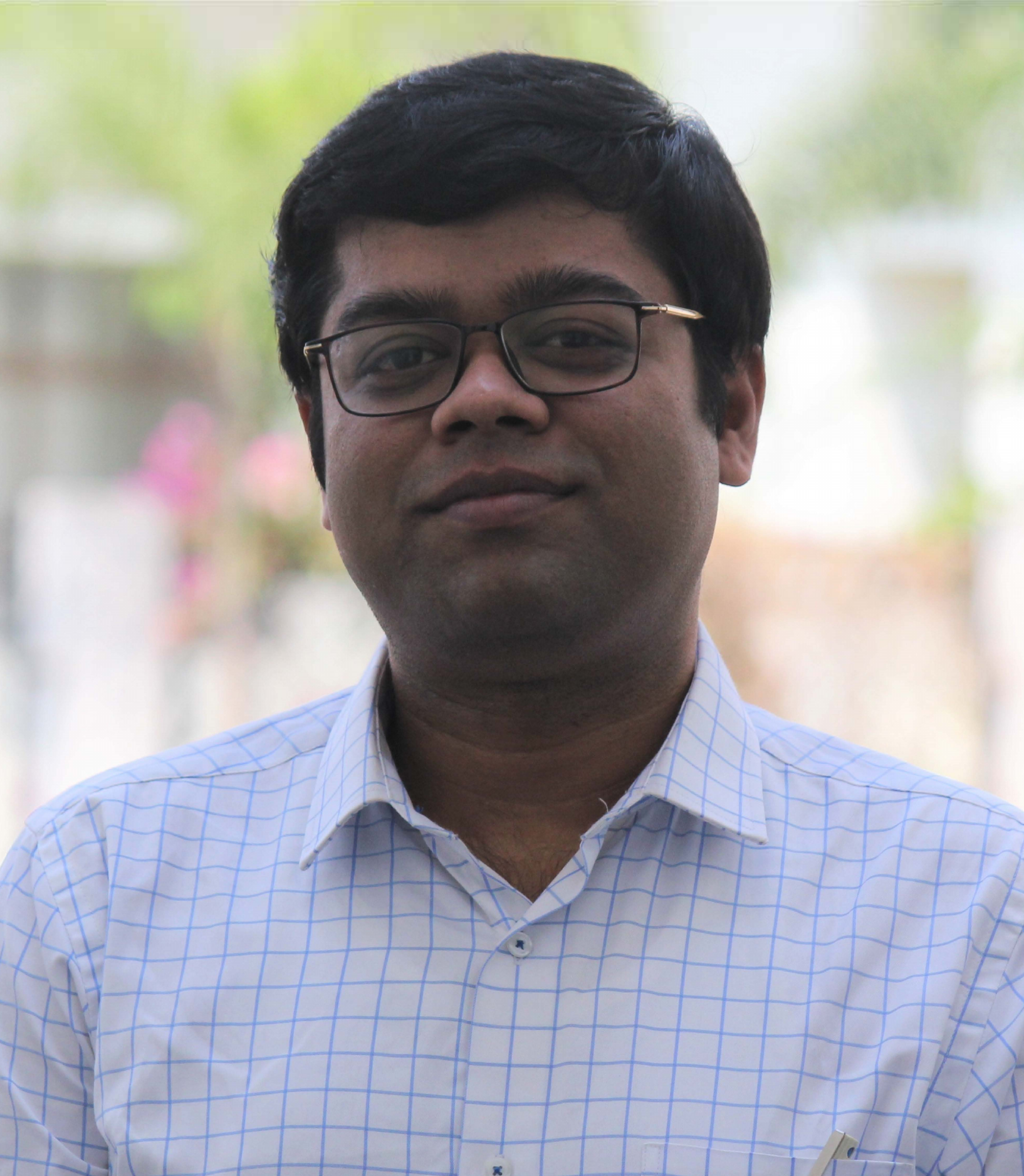}}
\noindent {\bf Debanjan Das} (Senior Member, IEEE) received the B.Tech. degree in applied Electronics and Instrumentation engineering from the Heritage Institute of Technology, Kolkata, in 2009, the M.Tech. degree in Instrumentation from Indian Institute of Technology, Kharagpur, in 2011, and the Ph.D. degree in Electrical Engineering from Indian Institute of Technology, Kharagpur, in 2016. He is an Assistant Professor with Dr. SPM IIIT Naya Raipur. His current research interests include IoT-Smart Sensing, Signal Processing, Bioimpedance, Instrumentation. He has been a member of the IEEE Engineering in Medicine and Biology Society, Measurement and Instrumentation Society.\par

%\subsection*{ }
\parpic{\includegraphics[width=1in,clip,keepaspectratio]{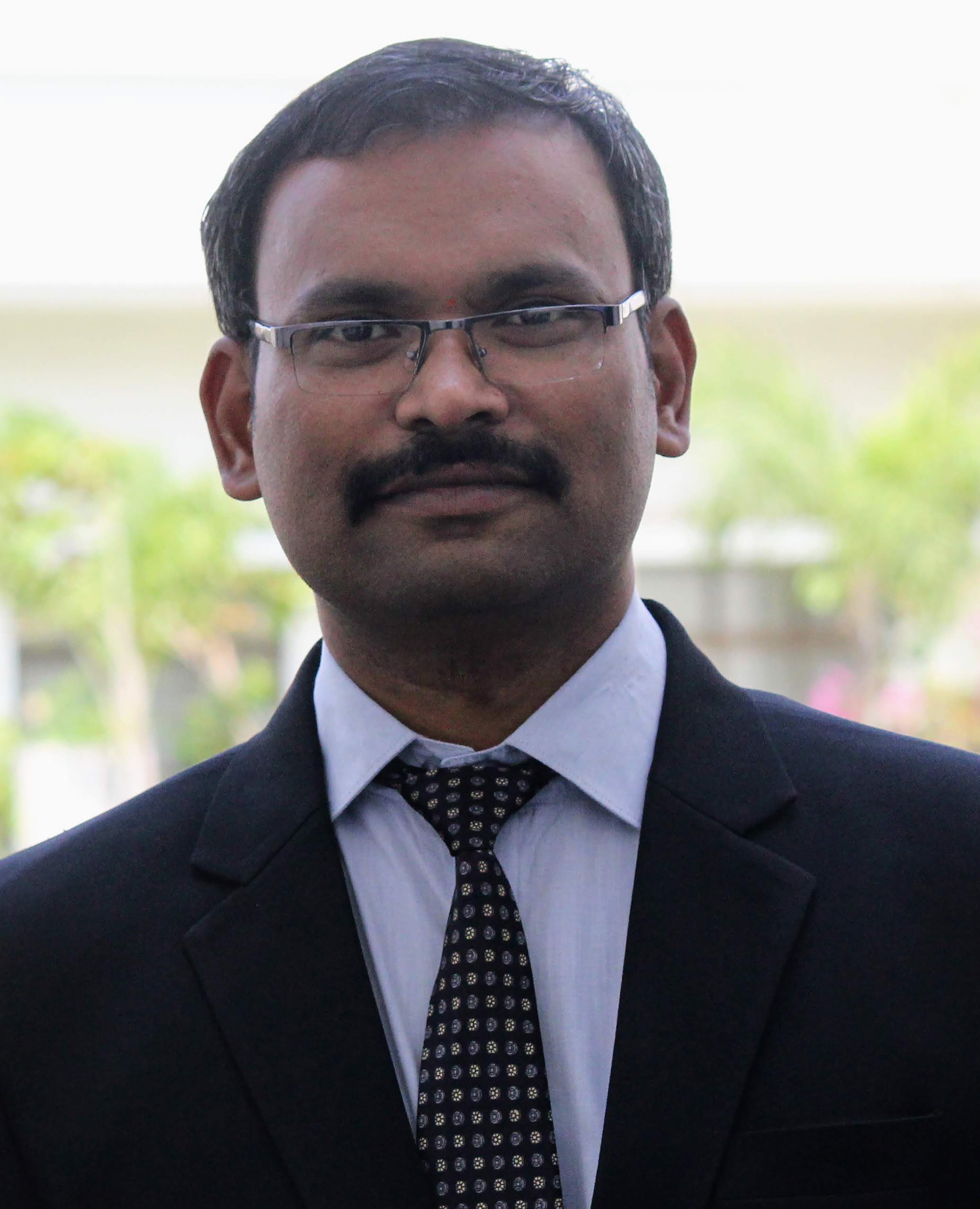}}
\noindent {\bf Venkanna Udutalapally} (M’15) obtained his Ph.D. degree by the National Institute of Technology, Tiruchirappalli (NITT), in 2015. Since 2005, he has been in the teaching profession and currently he is an Assistant Professor in the Department of CSE, IIIT Naya Raipur (IIIT- NR). He has eight years of teaching experience and seven years of research experience. His research interests include Internet of Things (IoT), Software Defined Networks, Network Security, Wireless Ad hoc, and Sensor network. He is a recipient of best paper award in IEEE-ANTS-2019 conference.\par

\subsection*{ }
\parpic{\includegraphics[width=1in,height=1.25in,clip]{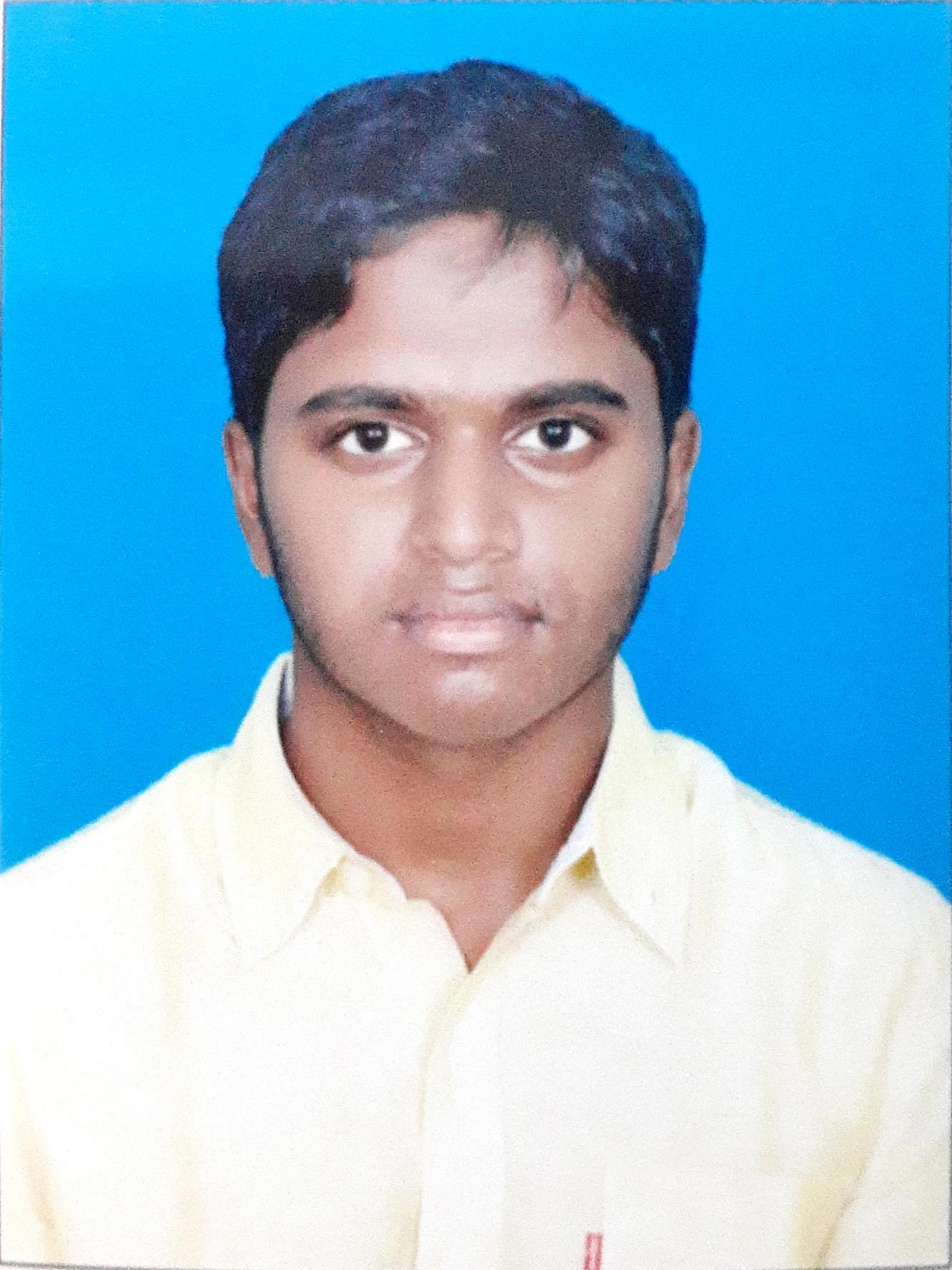}}
\noindent {\bf Mukil Kumar Selvarajan} is currently pursuing bachelor's degree in Instrumentation and Control Engineering from the National Institute of Technology, Tiruchirappalli (NITT).\par

\subsection*{ }
\subsection*{ }
\subsection*{ }

\parpic{\includegraphics[width=1in,clip,keepaspectratio]{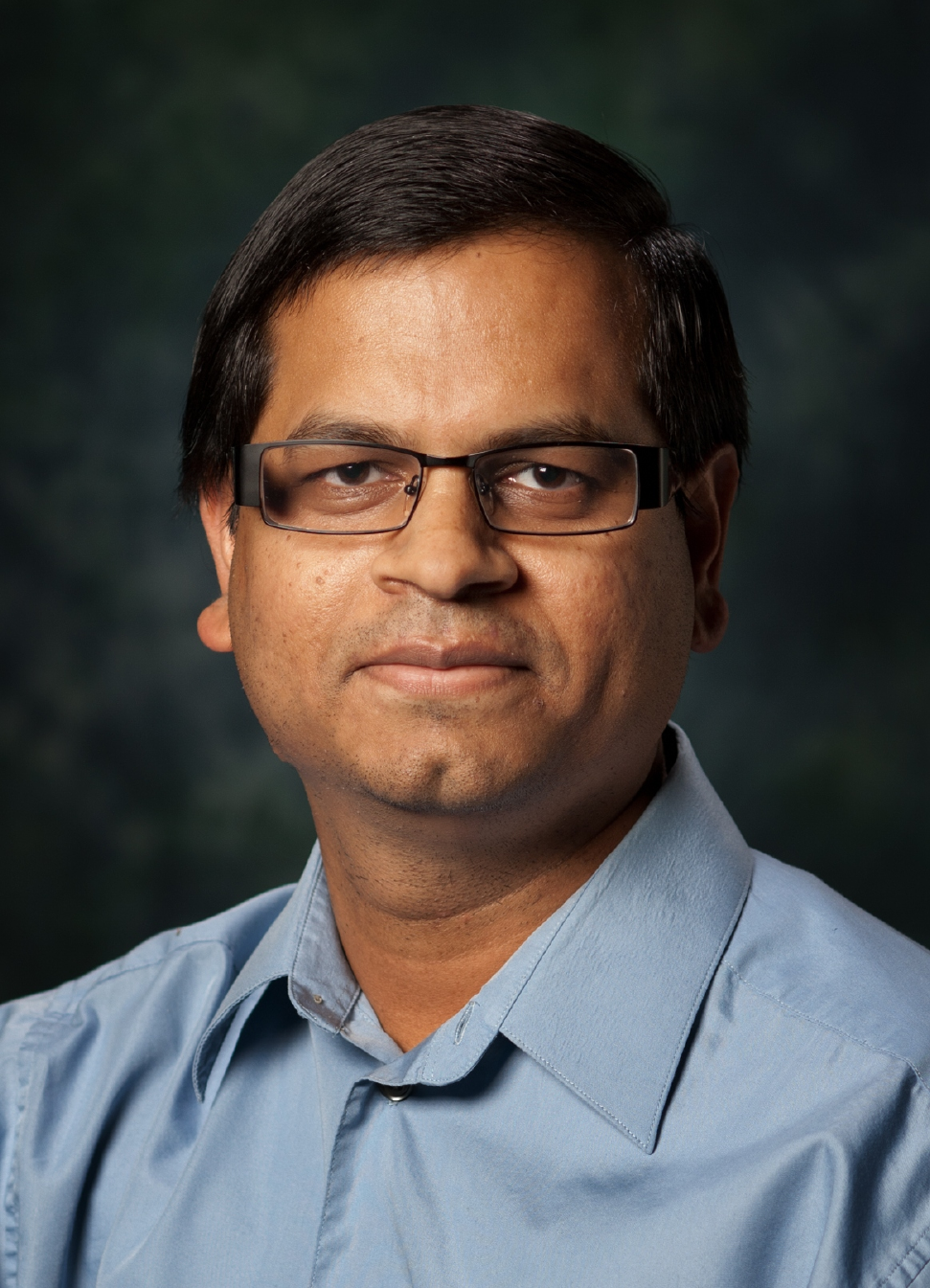}}
\noindent {\bf Saraju P. Mohanty} (Senior Member, IEEE) received the bachelor’s degree (Honors) in electrical engineering from the Orissa University of Agriculture and Technology, Bhubaneswar, in 1995, the master’s degree in Systems Science and Automation from the Indian Institute of Science, Bengaluru, in 1999, and the Ph.D. degree in Computer Science and Engineering from the University of South Florida, Tampa, in 2003. He is a Professor with the University of North Texas. His research is in ``Smart Electronic Systems'' which has been funded by National Science Foundations (NSF), Semiconductor Research Corporation (SRC), U.S. Air Force, IUSSTF, and Mission Innovation. He has authored 350 research articles, 4 books, and 7 granted and pending patents. His Google Scholar h-index is 42 and i10-index is 156 with 7400 citations. He is regarded as a visionary researcher on Smart Cities technology in which his research deals with security and energy aware, and AI/ML-integrated smart components. He introduced the Secure Digital Camera (SDC) in 2004 with built-in security features designed using Hardware-Assisted Security (HAS) or Security by Design (SbD) principle. He is widely credited as the designer for the first digital watermarking chip in 2004 and first the low-power digital watermarking chip in 2006. He is a recipient of 13 best paper awards, Fulbright Specialist Award in 2020, IEEE Consumer Technology Society Outstanding Service Award in 2020, the IEEE-CS-TCVLSI Distinguished Leadership Award in 2018, and the PROSE Award for Best Textbook in Physical Sciences and Mathematics category in 2016. He has delivered 11 keynotes and served on 12 panels at various International Conferences. He has been serving on the editorial board of several peer-reviewed international journals, including IEEE Transactions on Consumer Electronics (TCE), and IEEE Transactions on Big Data (TBD). He is the Editor-in-Chief (EiC) of the IEEE Consumer Electronics Magazine (MCE). He has been serving on the Board of Governors (BoG) of the IEEE Consumer Technology Society, and has served as the Chair of Technical Committee on Very Large Scale Integration (TCVLSI), IEEE Computer Society (IEEE-CS) during 2014-2018. He is the founding steering committee chair for the IEEE International Symposium on Smart Electronic Systems (iSES), steering committee vice-chair of the IEEE-CS Symposium on VLSI (ISVLSI), and steering committee vice-chair of the OITS International Conference on Information Technology (ICIT). He has mentored 2 post-doctoral researchers, and supervised 13 Ph.D. dissertations, 26 M.S. theses, and 11 undergraduate projects.

%\begin{wrapfigure}{l}{25mm} 
%    \includegraphics[width=1in,height=1.25in,clip,keepaspectratio]{AparnaSinha.png}
 % \end{wrapfigure}\par
%  \textbf{Aparna Sinha}  (Student Member, IEEE) received the B.Tech. degree in Electronics and Communication Engineering from Techno India Saltlake, Kolkata, in 2013, and the M.Tech. degree in VLSI Design from Department of Radiophysics and Electronics, University of Calcutta, in 2020. She is currently working towards the Ph.D. degree from IIIT Naya Raipur, India. 
%\\ Her current research interest includes Internet of Things (IoT) and Sensors.\par

\end{document}